\documentclass{aastex631}
\usepackage{graphicx}
\begin{document}

\title{DISTANCES TO 10 NEARBY GALAXIES OBSERVED
WITH THE HUBBLE SPACE TELESCOPE} 

\author{I.D.Karachentsev} 
\affil{Special Astrophysical Observatory, the Russian Academy of Sciences, Nizhnij Arkhyz, Karachai-Cherkessian Republic,
   Russia 369167}
   \email{ikar@sao.ru}

  \author{N.A.Tikhonov} 
  \affil{Special Astrophysical Observatory, the Russian Academy of Sciences, Nizhnij Arkhyz, Karachai-Cherkessian Republic,
   Russia 369167}
   \email{ntik@sao.ru}

\begin{abstract}

Images of 10 galaxies in F814W and F606W filters obtained on the Hubble Space Telescope (HST)
are used to construct color-magnitude diagrams for the star population of these galaxies. The
distances to the galaxies are estimated from the luminosity of the tip of the red giant branch. The
galaxies examined here have radial velocities from 250 to 760 km/s relative to the centroid of the
Local Group and distances ranging from 3.7 to 13.0 Mpc. Several other observed galaxies with low
radial velocities are noted at distances beyond the limit of 13 Mpc.
\end{abstract}


 \section{Introduction}
Considerable advances in the study of the kinematics and dynamics of the Local universe over the last two
decades have resulted from massive determination of highly accurate distances to nearby galaxies with the aid of the
Hubble Space Telescope (HST). Using the luminosity of the tip of the red giant branch (TRGB) (Lee et al. 1993) it has become
possible to measure distances with an error of $\sim5$\% for more than 400 galaxies in the Local volume (LV) within
distances of
D$\approx$11
Mpc around us. For many galaxies their membership in nearby groups has been established
and based on the virial velocities of the satellites the masses of the dark halo surrounding the bright central galaxies
 (Karachentsev and  Kashibadze 2021) has been determined. The average density of the dark matter concentrated in the halos of the local galaxies
turned out to be equal to 0.08 in units of the critical density (Karachentsev and Telikova 2018), which forms only a quarter of the global cosmic
density of matter $\Omega_m\approx0.31$.
This result seems paradoxical since the average density of stellar mater in the
Local volume is practically indistinguishable from the average global stellar density. The deviation of the radial
velocities of the galaxies from the ideal Hubble flow $V = H_0 D$, where $H 0 = 73$ km/s/Mpc is the Hubble parameter,
also provided precise evidence of the existence of a collective motion of the local galaxies toward a nearby rich
cluster in the Virgo constellation, as well as of their participation in the systematic expansion of the neighborhoods
of the Local void. Observational data on the distances of the galaxies in the LV presented in the Updated Nearby
Galaxy Catalog (UNGC) (Karachentsev et al. 2013) and Extragalactic Distance Database (EDD) (Anand et al. 2021) are the basis for testing cosmological
models on small-scales.

The latest significant supplement to the data on the distances of nearby galaxies were the observations on the
HST with the ACS camera in the SNAP 15922 program (PI, R. B. Tully). Of the 153 targets in the program,
observations were carried out for 80 galaxies. Accurate TRGB-distances were obtained for 53 galaxies 
(Anand et al 2021,  Karachentsev et al. 2020a, Karachentsev et al. 2020b, Karachentsev et al. 2021, Karachentsev et al. 2022a, Karachentsev et al. 2022b, Karachentsev et al. 2023) and
documented in the EDD data base (edd.ifa.hawaii.edu). Independent measurements of TRGB-distances for 24
dwarf galaxies from the SNAP15922 survey were made by Tikhonov and Galazutdinova (2022); for half of these
galaxies the distances were determined for the first time. Among the observed objects, cases remained in which the
guiding of the telescope was not ideal, along with galaxies with a TRGB position near the limit of the observations
and several blue compact galaxies with a high stellar concentration. The examination of these complex cases is the
subject of our paper.

\section{Observations and data processing}
Images of these galaxies were obtained with the ACS HST camera with exposures of 760 s in both the
F814W and F606W filters. The choice of exposure was determined by the feasibility of carrying out a cycle of
observations within a single orbital period of the HST in the SNAP-regime.
The standard program packages DAOPHOT II (Stetson 1987, Stetson 1994) and DOLPHOT 2.0 (Dolphin 2016) were used for stellar
photometry. The results of the photometry of the stars underwent customary testing in accord with the ``$\chi$'' and 
``SHARP'' criteria, in order to exclude diffuse objects. The details of the method are discussed in Tikhonov et al. (2019).
Images of the 10 galaxies that we have studied are shown as a mosaic in Fig. 1. The size of each image in the F606W filter is 1.67$^{\prime}$. Some large galaxies only fall partially into this format.

\begin{figure}
\includegraphics[height=9cm]{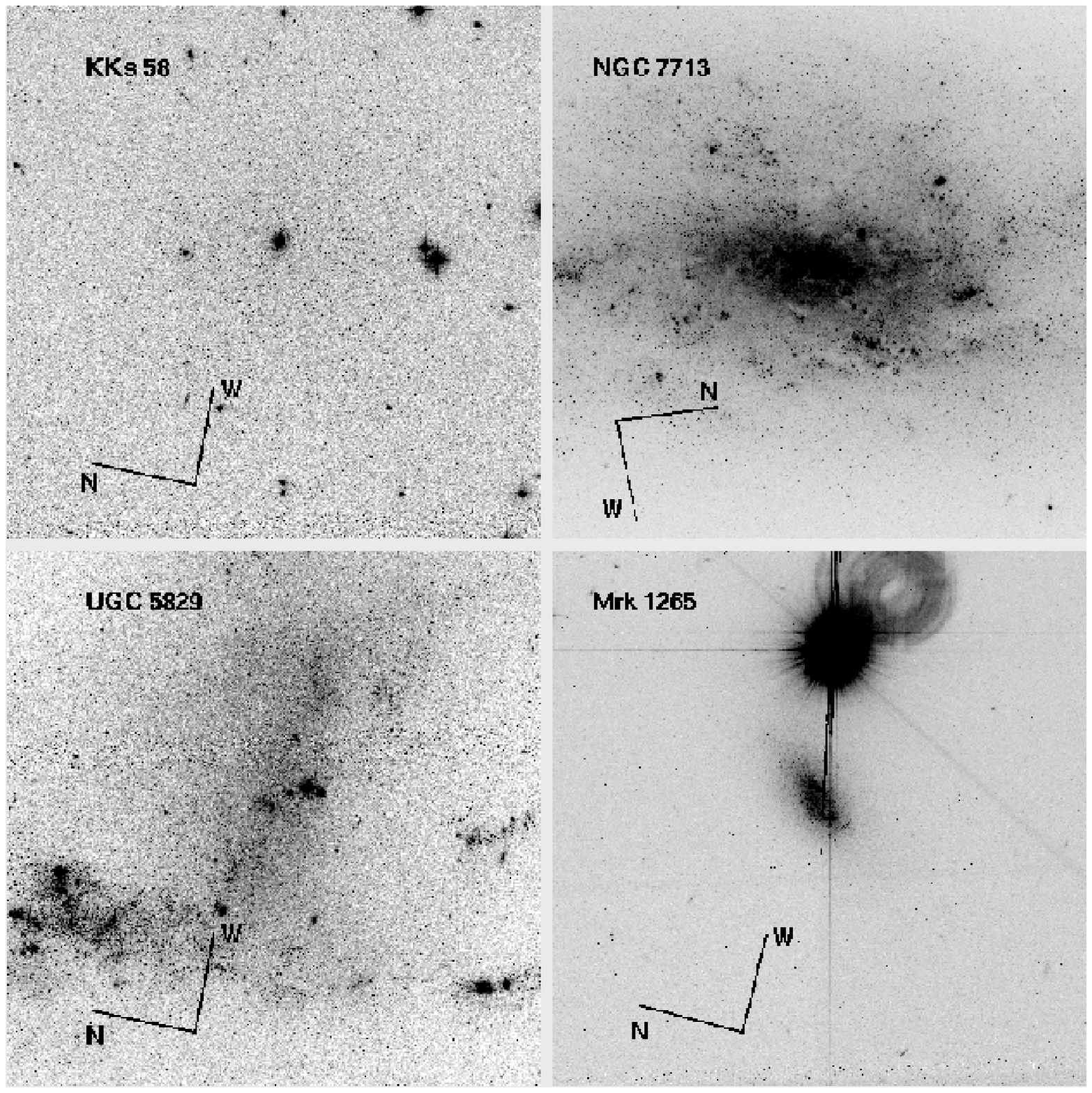}
\includegraphics[height=9cm]{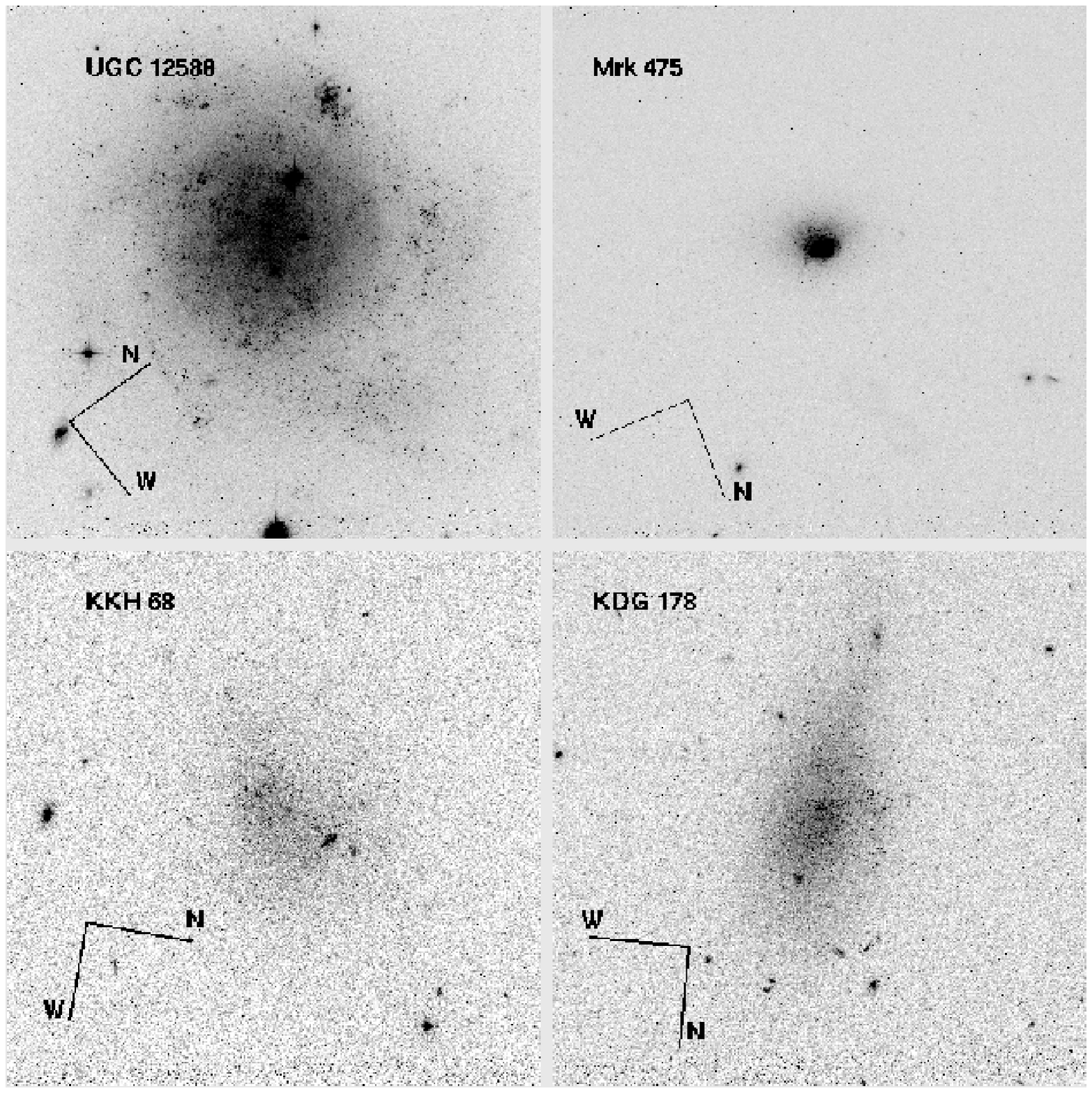}
\includegraphics[height=4.5cm]{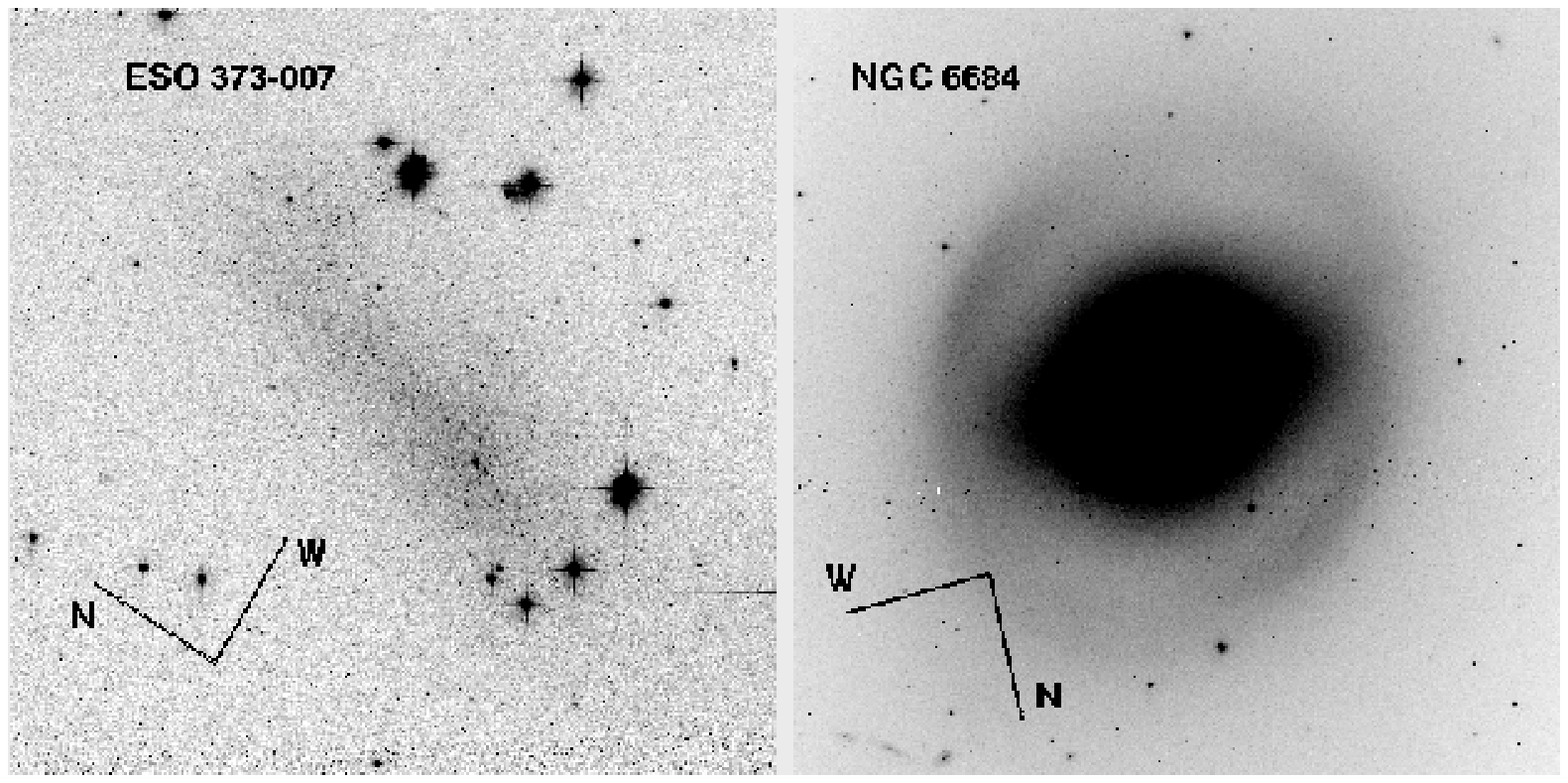}
\caption{Images of the galaxies in the F606W filter obtained on
the ACS camera of the Hubble Space Telescope. The size of the
pictures is $1.67^{\prime}\times1.67^{\prime}$; north and west are indicated by arrows.}
\end{figure}

The resulting color-magnitude diagrams (CMD) for each of the 10 galaxies are shown in the left-hand
frames of Fig. 2. The right-hand panels of the figure reproduce the constructed luminosity function of the galaxies
on a linear scale.
The luminosity function was constructed using stars with a color index in the range
$1.0 < (V-1) < 1.7$, corresponding to the range of color of the stars in the red giant branch. The position of the TRGB
of the galaxies is indicated by a horizontal line. It was determined from the jump in the luminosity function of the
stars, for which the Sobel function (Madore and Freedman 1995), the maxima of which correspond to sharp changes in the number of stars,
was used.
This method has known difficulties that have been repeatedly discussed in the literature. These
difficulties show up when the TRGB-galaxies lie near the photometric limit (in our case $I_{\rm lim}\approx27.0^m$) or if the
number of measured stars in a dwarf galaxy is low.
 
\begin{figure}
\includegraphics[height=14cm]{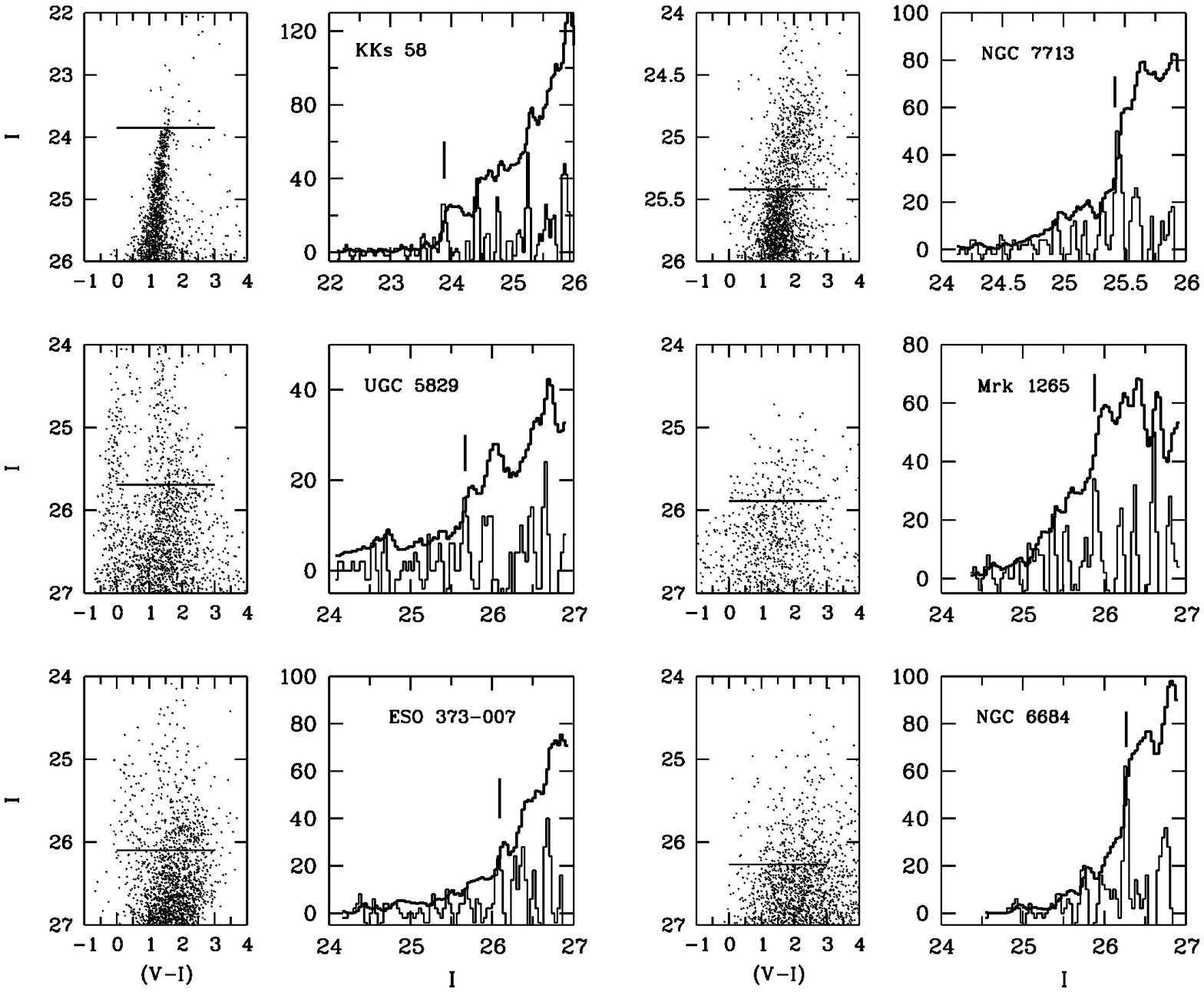}
\includegraphics[height=9.2cm]{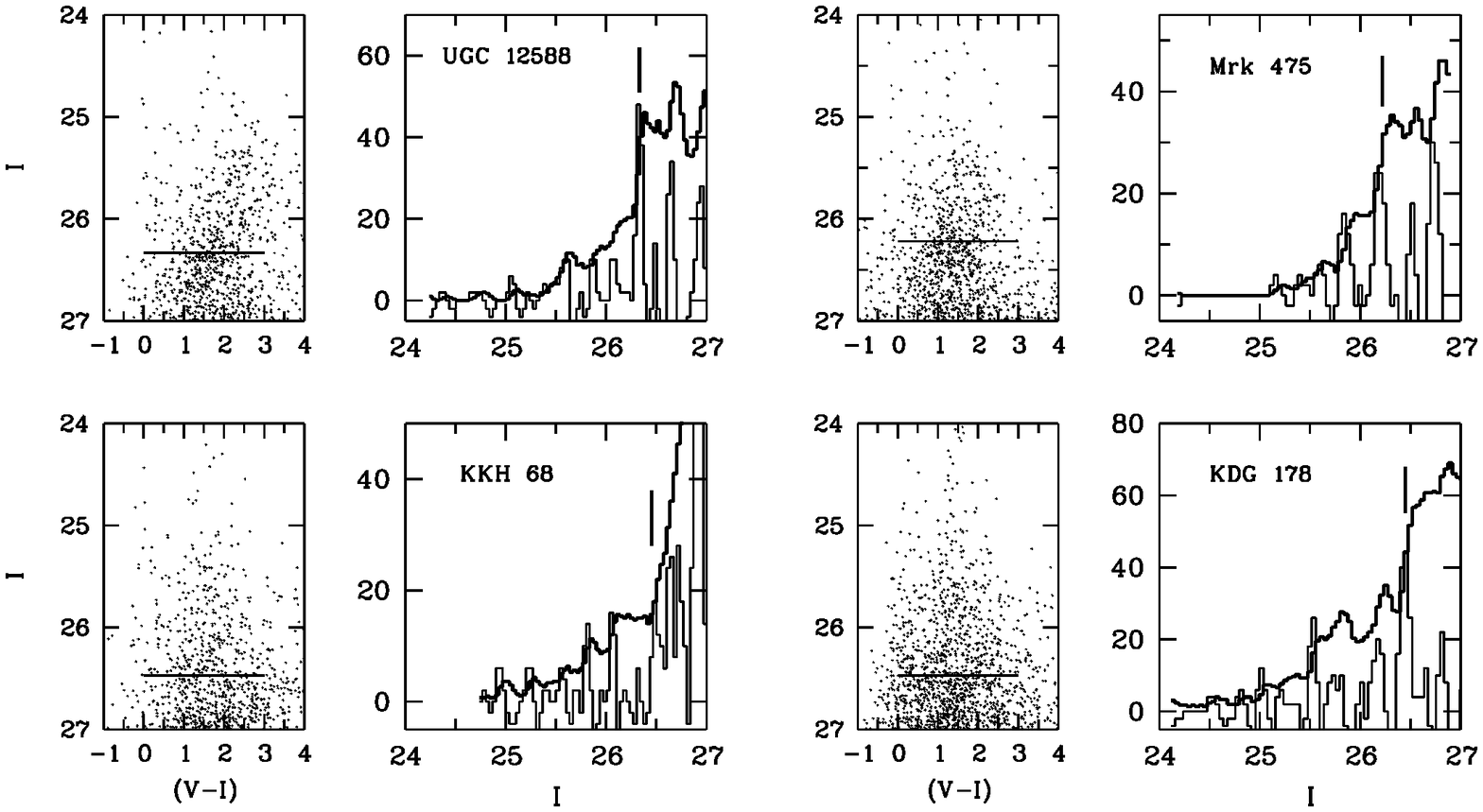}
\caption{CM diagrams for the galaxies examined here (left frames). The horizontal
lines indicate the position of the TRGB. The right hand panels show the luminosity
function of the red stars on a linear scale.}
\end{figure}

Yet another factor that makes it difficult to determine the position of the TRGB is the presence in a galaxy of
subsystem of AGB-stars with luminosities that are only slightly brighter than the luminosity of the tip of the red giant branch. In order to attenuate the role of this effect, for photometry we chose stars that are on the margins of
the galaxy, where the relative contribution of AGB-stars is smaller than the contribution of the RGB-population.
The measured distances for the galaxies examined here are listed in Table 1, where (1) is the name of the
galaxy in the sao.ru/lv/lvgdb data base; (2) is the number of the galaxy in the HyperLEDA catalog (Makarov et al 2014); (3) are the
equatorial coordinates in the epoch J2000.0; (4) is the radial velocity in km/s relative to the centroid of the Local
group; (5) is the position of the tip of the red giants branch (mag.); (6) is the interstellar absorption in the direction
of the galaxy (mag.) according to Schlafly and Finkbeiner (2011); (7) the position of the TRGB (mag.), which depends
weakly on the average color index (F814W--F606W) in the population of the RGB (Rizzi et al. 2007); (8,9) is the modulus of the
distance of a galaxy (mag.) and the linear distance in Mpc. The galaxies in the table are ranked in order of their estimated distances.
\begin{table}
 \caption{TRGB properties of the observed galaxies.}
 \begin{tabular}{c|c|c|c|c|c|c|c|c} \hline

   Galaxy      &           PGC &   RA(2000.0)Dec  & $V_{\rm LG}$   & $I_{\rm TRGB}$ &  $A_I$  &   $M_I$ & $DM$ &   $D$   \\ \hline

     (1)       &        (2)    &      (3)         &  (4)   &   (5)  &  (6)  &   (7)   &   (8)   &   (9) \\ \hline
  
 [KK2000]58    &      2815824  &  134600.8-361944 &  +255  &  23.85 &  0.09 &  $-$4.11  &  27.87  &  3.75 \\  
 NGC 7713      &        71866  &  233615.0-375620 &  +696  &  25.45 &  0.02 &  $-$4.10  &  29.53  &  8.05 \\
 UGC 5829      &        31923  &  104241.9+342656 &  +592  &  25.69 &  0.04 &  $-$4.11  &  29.76  &  8.95 \\
 Mrk 1265      &         32413 &  104940.4+225019 &  +534  &  25.89 &  0.04 &  $-$4.05  &  29.90  &  9.55 \\
 ESO 373-007   &      27104    &  093245.4-331444 &  +556  &  26.10 &  0.20 &  $-$4.05  &  29.95  &  9.77 \\
 NGC 6684      &       62453   &  184857.9-651024 &  +720  &  26.27 &  0.10 &  $-$3.88  &  30.05  & 10.23 \\
 UGC 12588     &      71368    &  232442.4+412053 &  +723  &  26.33 &  0.22 &  $-$4.11  &  30.22  & 11.07 \\
 Mrk 475       &         52358 &  143905.4+364822 &  +677  &  26.22 &  0.02 &  $-$4.11  &  30.31  & 11.53 \\
 KKH 68        &      2807141  &  113053.3+140846 &  +753  &  26.47 &  0.06 &  $-$4.07  &  30.48  & 12.47 \\
 KDG 178       &        42413  &  124010.0+323931 &  +763  &  26.47 &  0.02 &  $-$4.12  &  30.57  & 13.00 \\
\hline
\end{tabular}
\end{table}

 A comparison of the distances estimated in this way by different authors shows that the typical error in
 measuring $D$ is $\sim5$\%. This error rises to 7--10\% as the TRGB of the galaxy approaches the photometric limit. Wherein,
 there is sometimes also a systematic error owing to confusion in the position of the tip of the branches of the RGB
 and AGB stars.
 
 It is clear that the data obtained here on the estimates of the distances of the galaxies require more detailed
 commentary.
 
 \section{Individual cases}
 {\em [KK2000]58.} This dwarf spheroidal galaxy was first resolved into stars in images obtained with the
 earthbound 8-m VLT telescope (M\"{u}ller et al. 2019). The authors determined a distance to the galaxy of $3.36\pm0.11$ Mpc, which
 placed this galaxy on the leading edge of the group around Cen A (NGC 5128). Our estimate of the distance,
 $3.75\pm0.18$, corresponds better to the average distance of the group Cen A (3.68 Mpc). Near the center of the dSph-galaxy there is a globular stellar cluster, the radial velocity of which was measured by Fahrion et al. (2020). On the
images obtained from ACS HST the globular cluster is well resolved into stars of the red giant branch (Fig. 3). The
position of the TRGB in the cluster is consistent with the estimate of the TRGB over the entire body of the dwarf galaxy.

\begin{figure}
\includegraphics[height=9cm]{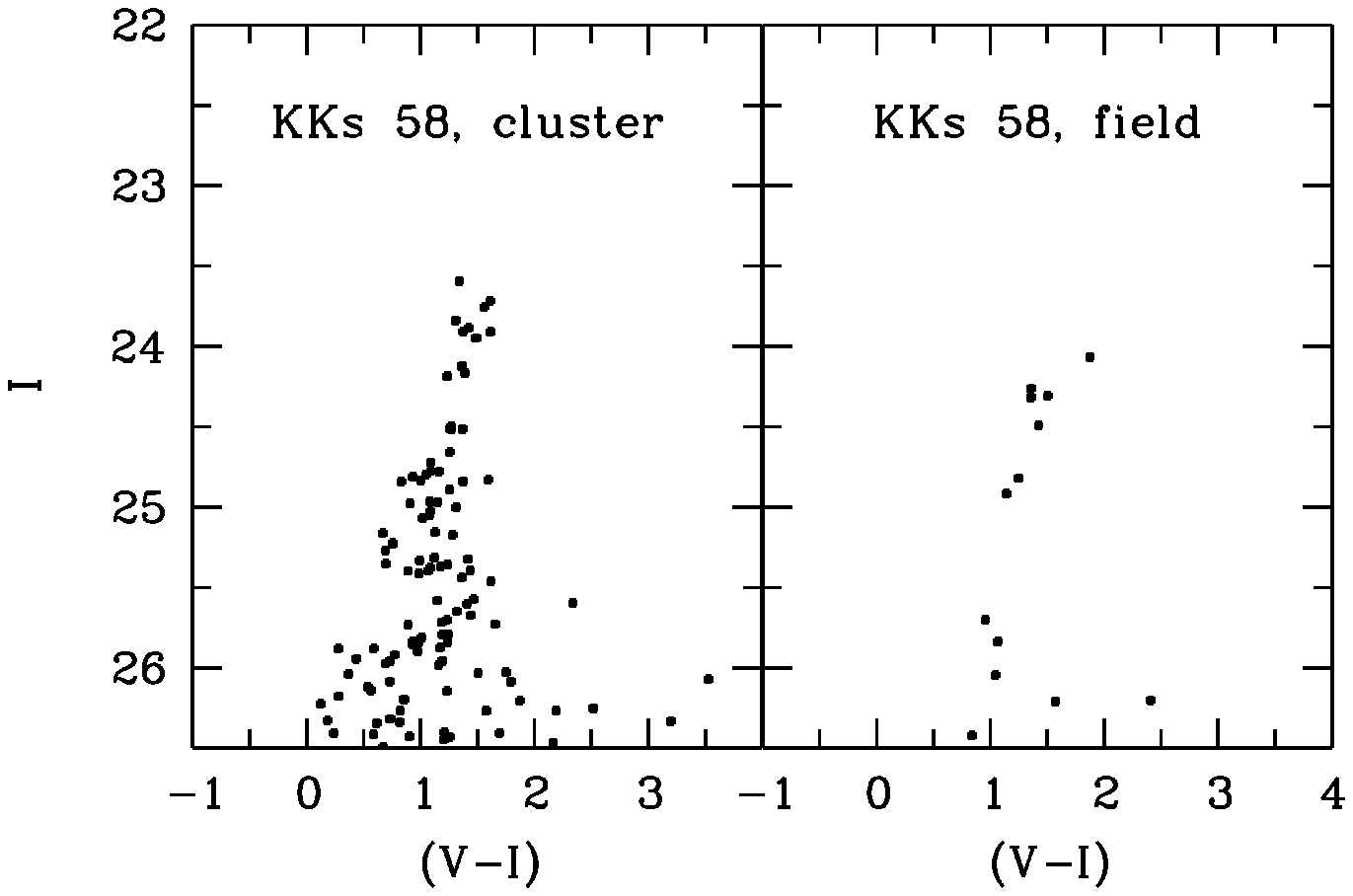}
\caption{Left: CM diagram for stars in the globular cluster near the center
of the dSph-galaxy [KK2000]58 in an aperture of radius $2^{\prime\prime}$. Right:
CM diagram for stars in the neighboring area with the same aperture.}
\end{figure}

{\em NGC 7713.} According to the NASA Extragalactic Database (NED; ned.ipac.caltech.edu), this spiral galaxy
of a late type has 23 estimates of distance by the Tully-Fisher method (Tully and Fisher 1977) which relate the luminosity of the galaxy
to the amplitude of its rotation. The average value of the distance from it is $9.05\pm1.38$ Mpc, which agrees, to within
the standard deviation, with our estimate of $8.05\pm0.40$ Mpc. At a projected distance of 1.9$^{\circ}$ from it (or 270 kpc)
there is another spiral galaxy of a late type, IC 5332, with a radial velocity $V_{\rm LG} = 716$ km/s and distance $D = 9.01$
Mpc (Anand et al 2021) that indicate a possible physical association of these galaxies.

{\em UGC5829=DDO 84=VV 794.} An irregular (Im) galaxy with a very peculiar structure that has
gotten the name ``spider'' (Fig. 4). More than half the baryon mass of the galaxy belongs to the gaseous component,
which evidently triggers numerous sources of star formation in it. UGC 5829 = KIG 434 shows up in the Catalog of
Isolated Galaxies (Karachentseva 1973). The nearest relatively massive galaxy NGC 3432 with $V_{\rm LG} = 578$ km/s and $D = 9.14$ Mpc (Anand et al 2021)
is at a projection distance of 3.0$^{\circ}$ or 470 kpc. UGC 5829 is a distinct example of how a distorted peculiar galactic
structure can be created, not by an external perturbation, but by strictly internal properties of a galaxy.

\begin{figure}
\includegraphics[height=9cm]{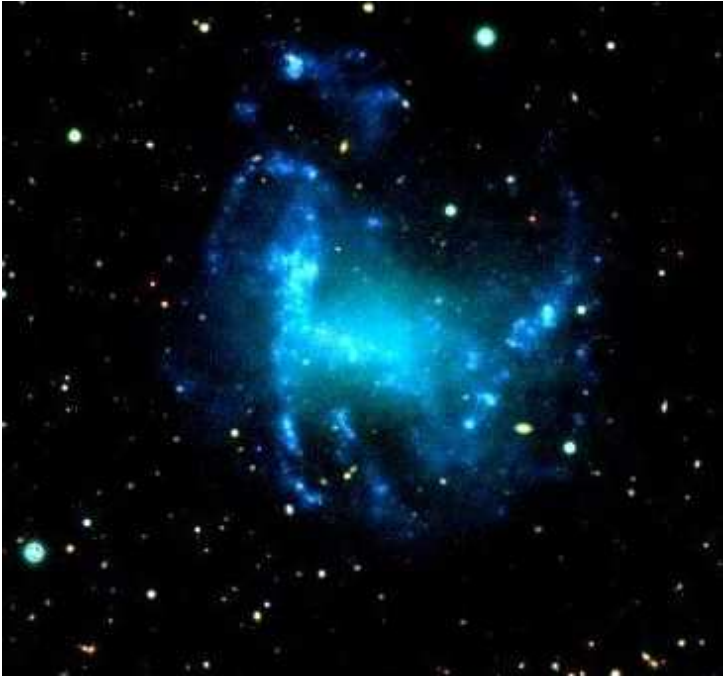}
\caption{An image of the peculiar galaxy UGC
5829 (VV 794, ``Spider''), taken from the digital
sky survey DECaLS. The size of the field is $6^{\prime}\times 6^{\prime}$,
north is upward, and east, to the left.}
\end{figure}

{\em Mrk 1265.} This blue compact  dwarf galaxy is in the aureole of a bright star in the Milky Way. The
compactness of the structure, abundance of blue stars, and presence of an aureole of the projected star lower the
accuracy of the photometry. According to our estimate, the distance to the galaxy is $D = 9.55\pm0.96$ Mpc, which is
somewhat greater than the distance estimate $D_{\rm NAM} = 7.8$ Mpc from the radial velocity taking into account the local
field of the velocities in the model of the Numerical Action Method (NAM, Shaya et al. 2017).

{\em ESO 373-007=AM 0930-330=[KK 2000]31.} The two estimates of the distance of this dIrr-galaxy, 8.32
Mpc according to Tully-Fisher and 9.65 Mpc in the NAM model are in good agreement with our estimate
$D = 9.77\pm0.98$ Mpc via  the TRGB. Nearby, at a projected distance of $15^{\prime}$ or 40 kpc there is a brighter
Sd-galaxy ESO 373-008 with a radial velocity of $V_{\rm LG} = 620$ km/s and a TF-distance of 9.68 Mpc. The two galaxies
evidently form an isolated pair.

{\em NGC 6684.} This galaxy is of an early type S0a, in which structure a bar and a ring can are seen.  A star
loop and radial jet can be seen on the periphery. The NED presents 14 crude estimates of the distance by various
approximate methods in a range from 5.5 to 15.5 Mpc. The TRGB distance that we have measured, $10.23\pm1.9$
Mpc, lies right in the middle of this interval. At a distance $2.6^{\circ}$ from NGC 6684 there is a spiral galaxy NGC 6744
with several satellites. Its TRGB-distance, 9.51 Mpc (Anand et al 2021) and radial velocity $V_{\rm LG} = 720$ km/s show that these galaxies
are probably members of a single diffuse association.

{\em UGC 12588.} This dIrr-galaxy is seen face-on and lies in a zone of substantial galactic extinction. Near it, at
a projected distance of $42^{\prime}$ there is a brighter spiral galaxy NGC 7640 with $V_{\rm LG} = 668$ km/s and a TRGB distance of
8.43 Mpc (Anand et al 2021). Our estimate of the TRGB-distance of UGC 12588 is $11.07\pm1.1$ Mpc, which is substantially greater
than the distance of the neighboring galaxy. To test the possible physical coupling of these galaxies it is necessary
to make deeper observations of UGC 12588.

{\em Mrk 475.} This blue compact galaxy lies far from the other galaxies with close radial velocities. The dense
stellar field in the main body of the galaxy lowers the accuracy of the performed photometry. Our estimate of the
TRGB-distance, $11.53\pm1.15$ Mpc, turned out to be considerably higher than the kinematic estimate 9.18 Mpc
(NAM).

{\em KKH 68=AGC 212837.} The position of the TRGB for this dIrr-galaxy is only $0.5^m$ higher than the photometric limit,
 which makes our estimate of the distance $D_{\rm TRBG} = 12.47\pm1.25$ Mpc unreliable. This galaxy lies at
an angular distance of $2.9^{\circ}$ from the spiral Sb-galaxy NGC 3627 (M 66) with TRGB distance of
$D_{\rm TRBG} = 11.12\pm0.56$ Mpc (Hoyt et al. 2019), which is the brightest member of the group in the Leo constellation. Probably KKH68 belongs to the peripheral members of this group.

{\em KDG 178=BTS 147.} This gas-rich dwarf galaxy lies only at  26$^{\prime}$ from the bright spiral galaxy NGC 4631,
for which $V_{\rm LG} = 581$ km/s and $D_{\rm TRBG} = 7.35\pm0.10$ Mpc (EDD). Judging from the distance $13.0\pm1.3$ Mpc, which we
obtained, KDG 178 lies behind the group of dwarfs around NGC 4631. The large radial distance of this object from the
group NGC 4631 is indirectly confirmed by relatively wide width of the 21-cm line, $W_{50} = 52$ km/s, which corresponds
to a TF-distance of more than 11 Mpc.

Besides the cases enumerated above, we also note several galaxies observed in the SNAP 15922 program
but which appear to be definitely distant. Jerjen, et al. (2000) estimated a distance of 4.29 Mpc for the galaxy
ESO 219-010 (PGC 44110)by the method of fluctuations in surface brightness.
This galaxy is essentially
unresolved into stars in the images taken with the ACS. In terms of texture, a  distance of this galaxy can be roughly
estimated as $\simeq 15$ Mpc.

Two dwarf galaxies in the profile of the Virgo cluster, EVCC 67 and UGC 7983 with radial velocities,
respectively, of 458 and 565 km/s, turned out to be unresolved in the images taken with ACS. They are evidently
members of the Virgo cluster at a  distance of 16.5 Mpc.

The isolated dIrr-galaxy KKH 46 (PGC 2807128) has a radial velocity $V_{\rm LG} = 409$ km/s and a width of the
21-cm line of $W_{50} = 25$ km/s. It is in the so-called ``Anomalous velocity zone'', (Tully 1988) where the galaxies with distances about
16 Mpc have large negative peculiar velocities of about --700 km/s.

\section{Concluding comments}
Using the images obtained with the ACS-camera on the Hubble Space Telescope in the F814W and F606W
filters, we have carried out PSF-photometry and constructed color-magnitude diagrams for the stellar population of 10
nearby galaxies. The positions of the tips of the red giant branches have been determined based on the jump in the
luminosity function of the red stars and these have been used to estimate the distances to the galaxies. Besides the
spiral galaxies NGC6684 and NGC 7713, the other objects are related to dIrr, dIm, BCD, and dSph dwarf galaxies.
The radial velocities of the galaxies relative to the centroid of the Local group lie within an interval from 250 to 770
km/s and the distances that we have determined lie within a range from 3.7 to 13.0 Mpc with a median of
10.0 Mpc. We also have noticed several galaxies with radial velocities $V_{\rm LG} <600$ km/s, which are scarcely resolved into stars
and lie at distances greater than 13 Mpc. These galaxies are either members of the Virgo cluster or lie within a zone of anomalously high negative peculiar velocities. The resulting data supplement the picture of the field of the
peculiar velocities of the galaxies in the Local volume, that is produced by nearby attractors with different dark halo
mass.

This work was based on observations taken with the NASA/ESA Hubble Space Telescope under contract
NAS5-26555. The HyperLEDA, NED, and EDD data bases have been used in this paper. This work was supported
by a grant from the Ministry of Science and Higher Education of the Russian Federation, No. 075-15-2022-262
(13.MNPMU.21.0003).

{\bf REFERENCES}

1. M. G. Lee, W. L. Freedman, and B. F. Madore, Astrophys. J. 417, 553 (1993).

2. I. D. Karachentsev and O. G. Kashibadze, Astron. Nachr. 342, 999 (2021).

3. I. D. Karachentsev and K. N. Telikova, Astron. Nachr. 339, 615 (2018).

4. I. D. Karachentsev, D. I. Makarov, and E. I. Kaisina, Astron. J. 145, 101 (2013).

5. G. S. Anand, L. Rizzi, R. B. Tully, et al., Astron. J. 162, 80 (2021).

6. I. D. Karachentsev, L. N. Makarova, R. B. Tully, et al., Astron. Astrophys. 638, 111 (2020a).

7. I. D. Karachentsev, L. N. Makarova, R. B. Tully, et al., Astron. Astrophys. 643, 124 (2020b).

8. I. D. Karachentsev, R. B. Tully, G. S. Anand, et al., Astron. J. 161, 205 (2021).

9. I. D. Karachentsev, J. M. Cannon, J. Fuson, et. al., Astron. J. 163, 51 (2022a).

10. I. D. Karachentsev, L. N. Makarova, G. S. Anand, et al., Astron. J. 163, 234 (2022b).

11. I. D. Karachentsev, L. N. Makarova, B. S. Koribalski, et al., Mon. Not. Roy. Astron. Soc. 518, 5823 ( 2023).

12. N. A. Tikhonov and O. A. Galazutdinova, Astrophys. Bull. 77, 430 (2022).

13. P. B. Stetson, Publ. Astron. Soc. Pacif. 99, 191 (1987).

14. P. B. Stetson, Publ. Astron. Soc. Pacif. 106, 250 (1994).

15. A. Dolphin, DOLPHOT: Stellar photometry, Astrophysics Source Code Library, record ascl:1608.013 (2016).

16. N. A. Tikhonov, O. A. Galazutdinova, and G. M. Karataeva, Astrophys. Bull. 74, 257 (2019).

17. B. F. Madore and W. L. Freedman, Astron. J. 109, 1645 (1995).

18. D. Makarov, P. Prugniel, N. Terekhova, et al., Astron. Astrophys. 570A, 13 (2014).

19. E. F. Schlafly and D. P. Finkbeiner, Astrophys. J. 737, 103 (2011).

20. L. Rizzi, R. B. Tully, D. I. Makarov, et al., Astrophys. J. 661, 815 (2007).

21. O. Müller, M. Rejkuba, M. Pawlowski, et al., Astron. Astrophys. 629A, 18 (2019).

22. K. Fahrion, O. M\"{u}ller, M. Rejkuba, et al., Astron. Astrophys. 634A, 53 (2020).

23. R. B. Tully and J. R. Fisher, Astron. Astrophys. 54, 661 (1977).

24. V. E. Karachentseva, Soobschenia SAO 8, 3 (1973).

25. E. J. Shaya, R. B. Tully, Yu. Hoffman, et al., Astrophys. J. 850, 207 (2017).

26. T. J. Hoyt, W. L. Freedman, B. F. Madore, et. al., Astrophys. J. 882, 150 (2019).

27. H. Jerjen, K. C. Freeman, and B. Binggeli, Astron. J. 119, 166 (2000).28. R. B. Tully, in: Large-Scale Motions in the Universe (Princeton: Princeton Univ. Press), p. 169 (1988).
\end{document}